\documentclass[conference,10pt]{IEEEtran}

\usepackage{algpseudocode}
\algtext*{EndWhile}
\algtext*{EndIf}
\algtext*{EndFor}
\algtext*{EndProcedure}
\usepackage{amsmath}
\usepackage{amssymb}
\usepackage{array}
\usepackage{algorithm}
\usepackage{bm}
\usepackage[bookmarks=false, hidelinks, breaklinks=true]{hyperref}
\usepackage[noadjust]{cite}
\usepackage[font=small]{caption}
\usepackage{color}
\usepackage{comment}
\usepackage{dblfloatfix}
\usepackage[inline]{enumitem}
\usepackage{epsfig}
\usepackage{etoolbox}
\usepackage{fancybox}
\usepackage{float}
\usepackage{fixltx2e}
\usepackage{graphicx}
\usepackage{listings}
\usepackage{mathrsfs}
\usepackage{multirow}
\usepackage{multicol}
\usepackage{pifont}
\usepackage{placeins}
\usepackage{rotating}
\usepackage{setspace}
\usepackage{soul}
\usepackage[hang, tight]{subfigure}
\usepackage{threeparttable}
\usepackage{times}
\usepackage{url}
\usepackage{upgreek}
\usepackage{verbatim}
\usepackage{wrapfig}
\usepackage[usenames,dvipsnames]{xcolor}
\usepackage[]{siunitx}
\usepackage[super]{nth}
\usepackage{tabularx}
\usepackage{supertabular}
\usepackage{tikz}
\usepackage{bbm}
\usepackage{blindtext}
\usepackage{makecell}
\usepackage{adjustbox}
\usepackage{afterpage}
\usepackage{breakcites}

\graphicspath{{./_fig/}}

\makeatletter
\renewcommand\footnoterule{%
  \kern-3\p@
  \hrule\@width0.4\columnwidth
  \kern2.6\p@}
  \makeatother

\usepackage[hang,flushmargin]{footmisc}

\begin{document}

\title{\huge{SpecLLM: Exploring Generation and Review of VLSI Design Specification with Large Language Model}}

\DeclareRobustCommand*{\IEEEauthorrefmark}[1]{%
\raisebox{0pt}[0pt][0pt]{\textsuperscript{\footnotesize\ensuremath{#1}}}}

\author{\IEEEauthorblockN{Mengming Li\IEEEauthorrefmark{1},
Wenji Fang\IEEEauthorrefmark{1,}\IEEEauthorrefmark{2}, Qijun Zhang\IEEEauthorrefmark{1} and
Zhiyao Xie\IEEEauthorrefmark{1}\IEEEauthorrefmark{*}}
\vspace{.1in}
\IEEEauthorblockA{\IEEEauthorrefmark{1}Hong Kong University of Science and Technology \\
\IEEEauthorrefmark{2}Hong Kong University of Science and Technology (Guangzhou) \\
\IEEEauthorrefmark{*}Coresponding Author: eezhiyao@ust.hk}}

\maketitle
\thispagestyle{plain}
\pagestyle{plain}

\begin{abstract}

The development of architecture specifications is an initial and fundamental stage of the integrated circuit (IC) design process. Traditionally, architecture specifications are crafted by experienced chip architects, a process that is not only time-consuming but also error-prone. Mistakes in these specifications may significantly affect subsequent stages of chip design. Despite the presence of advanced electronic design automation (EDA) tools, effective solutions to these specification-related challenges remain scarce. Since writing architecture specifications is naturally a natural language processing (NLP) task, this paper pioneers the automation of architecture specification development with the advanced capabilities of large language models (LLMs).


The absence of a clear and precise definition of architecture specifications poses the first challenge in our research. To address this, we propose a structured definition of architecture specifications, categorizing them into three distinct abstraction levels. 
Leveraging this definition, we create and release a specification dataset\footnote{The dataset will be available at https://github.com/hkust-zhiyao/SpecLLM} by methodically gathering 46 architecture specification documents from various public sources. Our clear definitions of architecture specifications, coupled with the dataset we have formed, lay a solid foundation for prospective research in employing LLMs for architecture specifications.

Leveraging our definition and dataset, we explore the application of LLMs in two key aspects of architecture specification development: (1) Generating architecture specifications, which includes both writing specifications from scratch and converting RTL code into detailed specifications. (2) Reviewing existing architecture specifications. We got promising results indicating that LLMs may revolutionize how these critical specification documents are developed in IC design nowadays. By reducing the effort required, LLMs open up new possibilities for efficiency and accuracy in this crucial aspect of chip design.

    
\end{abstract}

\section{Introduction}

\begin{figure}[t!]
\vspace{-.1in}
\centering
\includegraphics[width=0.97\linewidth]{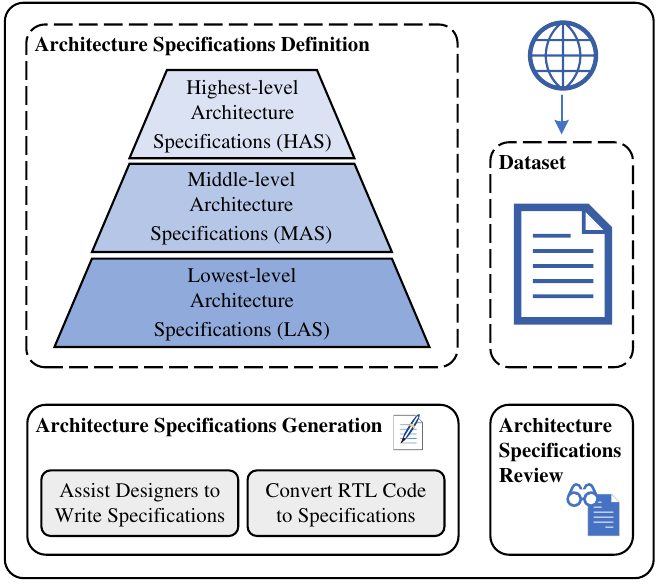}
\caption{The overall structure of this paper. We first propose basic definitions and an organized dataset dedicated to architecture specifications. Leveraging them, we explore the use of LLMs in the generation and review of architecture specifications.}
\vspace{-.1in}
\label{fig:framework}
\end{figure}

Developing architecture specifications is a critical initial step in the process of IC design. It lays the foundational framework and guidelines necessary for the subsequent stages of design and development. Traditionally, the task of writing and reviewing architecture specifications is undertaken by skilled chip architects. This process, while expertise-driven, tends to be time-consuming and can be susceptible to human errors, soliciting a more automated methodology to enhance efficiency and accuracy. 

In recent years, LLMs such as ChatGPT \cite{achiam2023gpt} have showcased remarkable capabilities in the field of artificial intelligence, with a wide range of applications from question answering to content creation. The growth of chip computing power will endow LLMs with greater capabilities. Consequently, researchers have started to investigate the potential of LLMs in augmenting the process of chip design, reversely enhancing the computing power of the chips themselves. For instance, recent studies~\cite{liu2023rtlcoder, blocklove2023chip, lu2023rtllm, liu2023verilogeval, thakur2023benchmarking, thakur2023autochip, nair2023generating, liu2023chipnemo} have utilized LLMs to generate RTL code like Verilog. Other works develop LLM-based solution to control EDA tools~\cite{liu2023chipnemo, he2023chateda}, design AI accelerator architectures~\cite{fu2023gpt4aigchip, yan2023viability}, hardware security assertion generation~\cite{kande2023llm}, fix security bugs~\cite{ahmad2023fixing}, etc. These research efforts imply a promising future for LLMs in chip design. In this paper, we conduct a pioneering investigation in the potential and practicality of LLMs in processing specifications. To the best of our knowledge, there has been no prior design automation or LLM research focusing on this important topic.


This paper focuses on employing LLMs to address the challenges inherent in the traditional management of architecture specifications. Figure~\ref{fig:framework} highlights the overall structure of this paper. Considering the absence of a formal definition or categorization of architecture specifications, we start with investigating existing architecture specifications across a diverse range of products. Then we categorize architecture specifications into three levels: Highest-level Architecture Specification (HAS), Middle-level Architecture Specification (MAS), and Lowest-level Architecture Specification (LAS). HAS is defined as the specification that establishes standards applicable across a range of products. It is at the highest abstraction level. MAS represents the specification that outlines the high-level architecture of a single-chip design. LAS is the detailed specification focused on the microarchitecture design of an individual chip.


Building upon our definition, we have assembled a comprehensive dataset that includes 46 public architecture specifications from various types of products. 
For the HAS, MAS, or LAS, the architecture
specifications related to RISC-V are among the most accessible and widely available. In contrast, architecture specifications for other instruction sets are relatively scarce. We have also observed that current architecture specifications suffer from a lack of unified writing standards, and the length of these specifications may exceed the processing capacity of LLMs. Both of them complicate the process of handling architecture specifications using LLMs.

Based on the basic formulation, we investigate the application of LLMs in both generation and reviewing architecture specifications. Regarding specification generation, we suggest two potential approaches. The first is to simplify the process of writing architecture specifications for designers, making it more efficient and less error-prone. The second approach is applicable when architecture specifications are absent for an already implemented chip. In such scenario, we can transform the RTL code back into architecture specifications, essentially reconstructing the original design documentation from the implemented code. According to our experiment, when generating specifications for simple logic circuits, the majority of the human tasks could be done by LLMs. We are optimistic that, even for more complex logic circuits, it is promising for LLMs to progressively take over a large portion of the work.





 
We also demonstrate that the LLMs can efficiently review the architecture specifications. We first propose our definition of various types of defects in architecture specification document. Building on this, we utilized these defects as the target responses and developed specific prompts, seeking for LLM's review. As for the public architecture specifications, our experiments have demonstrated that the LLMs could provide valuable feedback for enhancing these documents. Moving forward, we plan to extend our research to include the evaluation of the review results generated by the LLMs, streamlining the review process and enhancing the reliability of the outcomes.

In summary, we formulate the task of adopting LLMs in the development of architecture specifications. We provide fundamental definitions and explore applications of LLMs in the realm of architecture specification development, including the basic definitions of architecture specifications and various tasks that LLMs can perform in this context. Our key contributions are summarized below:



\begin{itemize}
    \item We provide structured definitions of architecture specifications, facilitating efficient utilization of LLMs in developing architecture specifications. (Section~\ref{sec:def})
    \item We generate a dataset of design specifications by methodically collecting and systematically organizing architecture specification documents from a variety of online sources. This paves the way for exploring LLMs in the development of architecture specifications. (Section~\ref{sec:dataset})
    \item We explore the use of LLMs as tools for generating architecture specifications, including assisting designers in writing these specifications and converting RTL code into comprehensive specifications. Our findings suggest that LLMs hold considerable promise in efficiently generating architecture specifications. (Section~\ref{sec:llmwrite})
    \item We explore the use of LLMs in the review of architecture specifications. We identify various potential defects that may arise in these specifications. Based on these identified defects, we have crafted specific processes and prompts to guide the LLMs in their review. Our experimental results indicate that LLMs are capable of providing valuable feedback for improving these documents in aspects of accuracy. (Section~\ref{sec:review})
\end{itemize}

\section{Our Definition on Architecture Specification} \label{sec:def}



\newcommand\mydot{\makebox[0pt][l]{$\bullet$}\hspace*{0.75em}}
\begin{table*}[p]
\small
\centering
\caption{Our proposed dataset for architecture specifications, including approximately 46 specification documents.} 
\label{tab:allSpec}
\vspace{.05in}
 \renewcommand{\arraystretch}{1.8}
\begin{tabular}{|p{0.09\linewidth}|p{0.27\linewidth}|p{0.27\linewidth}|p{0.3\linewidth}|} 
 \hline
 \textbf{Type} & \textbf{Highest-level Architecture \newline Specifications (HAS)} & \textbf{Middle-level Architecture \newline Specifications (MAS)} & \textbf{Lowest-level Architecture \newline Specifications (LAS)} \\  \hline
 CPU
 & \mydot RISC-V ISA Specifications, Unprivileged Specification \cite{URL:risc-v_isa_unpri} \newline
 \mydot RISC-V ISA Specifications, Privileged Specification \cite{URL:risc-v_isa_pri} \newline
 \mydot ARMv8-M Architecture Reference Manual \cite{URL:armv8-m} \newline
 \mydot Intel 64 and IA-32 Architectures Software Developer’s Manual \cite{URL:intelarchsoft} \newline
 \mydot The SPARC Architecture Manual, Version 9 \cite{sparcv9} \newline
 \mydot OpenRISC 1000 Architecture Manual \cite{URL:openrisc1000}
 
 & \mydot *The NEORV32 RISC-V Processor: Datasheet \cite{URL:NEORV32} \newline
 \mydot *OpenSPARC T1 Microarchitecture Specification \cite{URL:OpenSPARCT1} \newline
 \mydot *OpenSPARC T2 Core Microarchitecture Specification \cite{URL:OpenSPARCT2} \newline
 \mydot E31 Core Complex Manual \cite{URL:E31} \newline
 \mydot E51 Core Complex Manual \cite{URL:E51} \newline
 \mydot Arm Cortex‑A78 Core Technical Reference Manual \cite{URL:arm_CortexA78} \newline
 \mydot Arm Cortex‑X2 Core Technical Reference Manual \cite{URL:arm_CortexX2} \newline
 \mydot Arm Neoverse-N2 Core Technical Reference Manual \cite{URL:armn2} \newline
 \mydot Intel 64 and IA-32 Architectures Optimization Reference \cite{URL:intelarchopt} \newline
 \mydot OpenRISC 1200 IP Core Specification \cite{URL:openrisc1200}

 & \mydot The NEORV32 RISC-V Processor: Datasheet \cite{URL:NEORV32} \newline
 \mydot OpenSPARC T1 Microarchitecture Specification \cite{URL:OpenSPARCT1} \newline
 \mydot OpenSPARC T2 Core Microarchitecture Specification \cite{URL:OpenSPARCT2} \newline
 \mydot Amber 2 Core Specification \cite{URL:amber} \newline
 \mydot LXP32, a lightweight open source 32-bit CPU core, Technical Reference Manual \cite{URL:lxp32} \newline
 \mydot OpenMSP430, Texas Instruments \cite{URL:openmsp430} \newline
 \mydot NEO430, based on the Texas Instruments MSP430(TM) ISA \cite{URL:neo430}
  \\ \hline

 SoC
 & \mydot Efficient Trace for RISC-V \cite{URL:riscv_trace} \newline
 \mydot RISC-V External Debug Support \cite{URL:riscv_debug} \newline
 \mydot RISC-V IOMMU Architecture Specification \cite{URL:riscv_iommu} \newline
 \mydot RISC-V Advanced Interrupt Architecture \cite{URL:riscv_interrupt} \newline
 \mydot RISC-V Platform-Level Interrupt Controller Specification \cite{URL:riscv_interrupt2}
 & \mydot Freedom E310-G000 Manual \cite{URL:E310} \newline
 \mydot Freedom U540-C000 Manual \cite{URL:E540}
 & \mydot \#The NEORV32 RISC-V Processor: Datasheet \cite{URL:NEORV32} \newline
 \mydot OpenSPARC T2 System-On-Chip (SoC) Microarchitecture Specification \cite{URL:OpenSPARCT2SOC}
 \\ \hline

 Accelerator
 & \mydot RISC-V "V" Vector Extension \cite{URL:rvv} \newline
 \mydot Intel Advanced Performance Extensions (Intel APX) Architecture Specification \cite{URL:intelapx} \newline
 \mydot Intel Advanced Vector Extensions 10 (Intel AVX10) Architecture Specification \cite{URL:intelavx}
 &
 & \mydot NVIDIA Deep Learning Accelerator (NVDLA), Hardware Architectural Specification \cite{URL:nvdla} \\ \hline

 Bus \& \newline
 Network
 & \mydot TileLink Specification \cite{URL:tilelink} \newline
 \mydot AMBA5 CHI Architecture Specification \cite{URL:amba5chi} \newline
 \mydot AMBA5 ACE Protocol Specification (superseded by CHI) \cite{URL:amba5ace}
 \newline
 \mydot AMBA5 AXI Protocol Specification \cite{URL:amba5axi} \newline
 \mydot AMBA4 AXI and ACE Protocol Specification \cite{URL:amba4axi}
 &
 & \mydot 10GE MAC Core Specification \cite{URL:mac} \newline
 \mydot Ethernet IP Core Specification \cite{URL:eip} \newline
 \mydot I2C-Master Core Specification \cite{URL:i2c} \newline
 \mydot UART to Bus Core Specifications \cite{URL:uart} \newline \\ \hline
 
 Arithmetic 
 & 
 &
 & \mydot Elliptic Curve Group Core Specification \cite{URL:ecg} \newline
 \mydot Tate Bilinear Pairing Core Specification \cite{URL:pairing} \newline
 \mydot Tiny Tate Bilinear Pairing Core Specification \cite{URL:tiny_tate_bilinear_pairing}
 \\ \hline

 Crypto
 &
 &
 & \mydot AES Core Specification \cite{URL:aes} \newline
 \mydot SHA3 Core Specification \cite{URL:sha3}
 \\ \hline
\end{tabular}
\begin{tablenotes} \footnotesize
\item$^\star$Notes: *NEORV32 \cite{URL:NEORV32}, OpenSPARC T1 \cite{URL:OpenSPARCT1}, OpenSPARC T2 \cite{URL:OpenSPARCT2} include chapters meeting the criterion of MAS. \#NEORV32 \cite{URL:NEORV32} encompasses chapters describing the SoC implementation.
\end{tablenotes}
\end{table*}


We define the architecture specification as the document that describes the chip architecture prior to RTL coding. Writing architecture specifications for the target chip is usually the starting point of the IC design flow. The term \textbf{architecture specification} is specifically chosen to emphasize that the document captures the architectural aspects of a chip, distinguishing it from general specifications. We categorize the architecture specifications into three levels. 

\begin{itemize}
    \item \textbf{The Highest-level Architecture Specification (HAS)} establishes standards applicable to a range of products. A notable example is the RISC-V specifications \cite{URL:risc-v_spec}. It defines the ISA specifications (e.g., instruction formats, register usages) and Non-ISA specifications (e.g., trace). Designing specific RISC-V chips should comply with these specifications.
    \item \textbf{The Middle-level Architecture Specification (MAS)} outlines the high-level architecture of a single product. These specifications encompass the essential information required to profile the chip design. For example, one MAS for a RISC-V CPU may include an overview of the microarchitectures (e.g., block diagram) and their primary parameters (e.g., cache size).
    \item \textbf{The Lowest-level Architecture Specification (LAS)} details the microarchitecture design of a single product. Unlike HAS and MAS, LAS should give the implementation details for each microarchitecture, which may involve the ports, internal signals, pipelines, and the associated descriptions. By reading them, the designers are expected to write the corresponding RTL code correctly.
\end{itemize}

The architecture specifications are distinct from the user manuals. We define the user manuals as the documents that succinctly outline the chip design post-production. In contrast to architecture specifications, which primarily cater to chip designers, user manuals are typically crafted for the end users and programmers. Nonetheless, we note that certain contents in the user manuals exhibit similarities to the HAS or MAS. For example, as described in Section~\ref{sec:dataset}, we identify the \textit{ARMv8-M Architecture Reference Manual} \cite{URL:armv8-m}, \textit{Intel 64 and IA-32 Architectures Software Developer’s Manual} \cite{URL:intelarchsoft} as HAS; and a series of \textit{Arm Core Technical Reference Manual}s \cite{URL:armn2,URL:arm_CortexX2,URL:arm_CortexA78}, \textit{Intel 64 and IA-32 Architectures Optimization Reference} \cite{URL:intelarchopt} as MAS.

\section{Dataset Overview}
\label{sec:dataset}


Table~\ref{tab:allSpec} provides an overview of our proposed dataset for architecture specifications. It shows a variety of public architecture specifications and user manuals, organized by product types like CPU, SoC, Accelerator, Bus and Network, Arithmetic and Crypto, and by levels such as HAS, MAS, and LAS. It is important to note that in Table~\ref{tab:allSpec}, we only display a representative selection of about 50 architecture specifications. For documents exhibiting a high degree of similarity, like additional \textit{Arm Core Technical Reference Manuals} and \textit{AMBA Protocol Specifications}, we have omitted them for simplicity. Our collection has yielded several notable findings.

\textbf{Availability.} For the HAS, MAS, or LAS, the architecture specifications related to RISC-V are among the most accessible and widely available. Especially for the HAS, there are various established standards that exist, catering to different kinds of products, such as the CPU \cite{URL:risc-v_isa_pri, URL:risc-v_isa_unpri}, SoC \cite{URL:riscv_trace, URL:riscv_iommu, URL:riscv_interrupt, URL:riscv_interrupt2}, Accelerator \cite{URL:rvv} and Cache Coherence Bus \cite{URL:tilelink}. An interesting observation is that despite the vast number of open-source designs in the RISC-V ecosystem, there is a scarcity of available MAS and LAS documents. Many documents, such as the Boom Core \cite{celio2015berkeley} and XiangShan Core \cite{xu2022towards}, which describe the design of open-source chips, do not fulfill the standards required for formal architecture specifications. Therefore, we have chosen not to include these in our table. This gap highlights a potential area for improvement in the documentation and standardization of open-source chip designs. For the Arm and X86 ecosystem, the availability of formal architecture specifications is weaker. We can only find some reference manuals intended for programmers or end-users. Despite this limitation, upon investigation, it is observed that the contents of these manuals partially align with what one would expect in formal architecture specifications. Consequently, we have chosen to include these reference manuals for conducting research.

\textbf{Standard.} While writing architecture specifications is a common practice in commercial chip manufacturing, we observe the available open specification documents exhibit a variety of writing styles and lack unified writing standards. Within individual semiconductor companies, there's a tendency to maintain a consistent format for architecture specifications. Yet, the absence of unified writing standards across the industry leads to a diversity in the presentation and structuring of these documents, especially evident in open-source chip design \cite{URL:NEORV32, URL:amber, URL:lxp32, URL:openmsp430, URL:neo430}. This variation in writing styles and lack of standardized formats in architecture specifications pose additional challenges for LLMs in handling and accurately interpreting these documents.

\textbf{Length.} The length of HAS spans from just over 100 pages to more than 1000 pages. The lengths of MAS and LAS can vary widely, with their extent directly correlating to the complexity of the chip designs they describe. MASs outline the chip designs, and thus are usually shorter than the LASs, ranging from several pages to in excess of 100 pages. LASs extend from over 10 pages to hundreds of pages. The longer the architecture specifications are, the more contextual information the LLMs tend to process. However, current commercial LLM products have a limitation on the number of tokens they can handle, which is not infinite. In Section~\ref{sec:review}, as we will discuss, the increase in contextual information due to longer architecture specifications poses challenges. It can lead to a decrease in the accuracy of the LLM's output. Therefore, the length of the architecture specifications must be carefully considered when employing an LLM to manage these documents.

In the following sections, we will choose portions of them to demonstrate LLM's capabilities in developing architecture specifications. Constrained by the scarcity of existing architecture specifications, we consider certain sections of user manuals to function as architecture specifications.

\section{LLM Generates Architecture Specification}
\label{sec:llmwrite}

\subsection{Motivation}
\label{subsec:writemoti}

Traditionally, writing architecture specifications is a non-trivial, but critical step in the process of IC design. Our research indicates that chip architects spend a considerable amount of time writing architecture specifications for several reasons:

\begin{itemize}
    \item \textbf{Massive Information Organization:} Whether for HAS, MAS, or LAS, a significant amount of information must be organized. Systematically arranging them is crucial for the ensuing RTL coding process. While criteria for writing architecture specifications exist in separate companies, the complexity of this task escalates with the increasing scale of the designs.
    \item \textbf{Diverse Modules Interconnection:} MAS is responsible for establishing the interconnection between various hardware modules. This plays a critical role in ensuring the overall correctness of the complete product. Despite this, the process of integrating these modules often encounters challenges, such as the time-consuming tasks of discerning different interfaces and identifying the functions of each module.
    \item \textbf{Complicate Algorithm Implementation:} LAS delineates the structures and behaviors for each microarchitectural design. According to our observation, LAS is closest to the final RTL design among the three types of architecture specifications. When dealing with complex microarchitectural algorithms, such as branch prediction, writing LAS presents significant challenges, which include accurately defining signals and state machines.
\end{itemize}

LLMs are promising in solving the above challenges and assisting designers in writing architecture specifications. According to Section~\ref{sec:def}, HASs set the standards for a variety of products and are inherently determined by human decisions. Consequently, they are highly flexible and challenging to generate. 
We thus mainly focus on using LLMs to automatically generate the MAS and LAS. We explore two LLM-based applications: 

\begin{enumerate}
    \item LLMs can reduce human effort by writing architecture specifications (Section~\ref{subsec:assist}).
    \item Many open-source chip designs lack comprehensive and formal architecture specifications. In situations where these specifications are absent, LLMs could reversely convert the RTL code into detailed specification documents. This reverse engineering process can bridge the gap in the documentation and enhance understanding and accessibility of the chip designs (Section~\ref{subsec:convert}).
\end{enumerate}

In the upcoming sections, we will leverage the widely-used commercial LLM product---GPT-4 \cite{achiam2023gpt}, to demonstrate how this technology can be effectively exploited to enhance the process of writing architecture specifications.

\subsection{Assist Designers to Write Architecture Specifications}
\label{subsec:assist}

\begin{figure}[t]
\centering
\includegraphics[width=\linewidth]{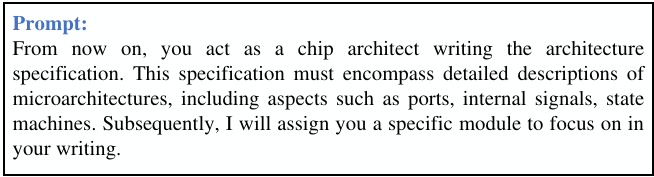}
\caption{An example prompt to initiate generating LAS. This generation is based on the designers' brief description.}
\vspace{-.1in}
\label{fig:startWrite}
\end{figure}

\begin{figure}[t]
\centering
\includegraphics[width=\linewidth]{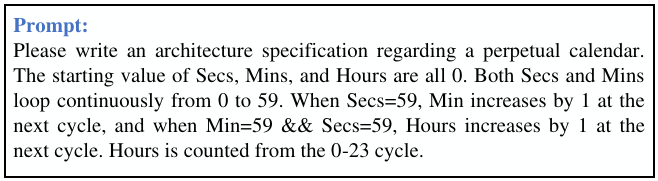}
\caption{An example prompt to request GPT generating an architecture specification for perpetual calendar. The response is shown in Figure~\ref{fig:writeResp}.}
\vspace{-.2in}
\label{fig:detailedWrite}
\end{figure}

\begin{figure}[p]
\centering
\includegraphics[width=\linewidth]{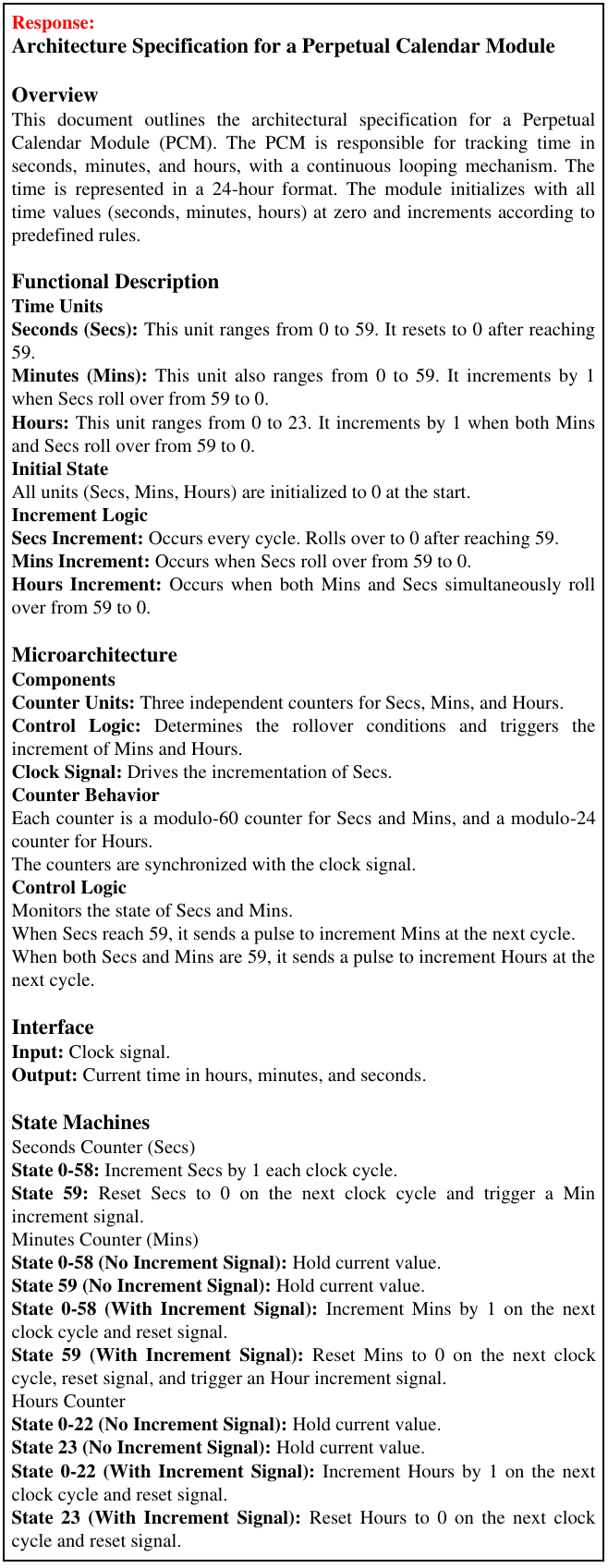}
\caption{A GPT response example to the prompt in Figure~\ref{fig:detailedWrite}.}
\vspace{-.1in}
\label{fig:writeResp}
\end{figure}

Figure~\ref{fig:startWrite} presents an example of a prompt to initiate the generation of LAS. This example, referencing \cite{white2023prompt}, indicates the specific role expected to be performed by the LLMs, and the expected response format. Then we employ the prompt in Figure~\ref{fig:detailedWrite} to direct GPT generating the architecture specifications. This particular example, sourced from \cite{lu2023rtllm}, has been selected due to its moderate level of complexity, making it suitable for testing purposes. Figure~\ref{fig:writeResp} presents the outcomes of GPT-4. The experimental results demonstrate that the GPT can accurately understand the questions. These specifications, spanning from the high-level functional description to the low-level state machines, offer in-depth information for subsequent RTL coding. Yet, the design of a perpetual calendar is a classical task in the field of circuit design. The LLMs have abundant corpus about it. The commercial products, usually include internal IPs, which are not used to train the external LLMs like ChatGPT. We believe it is non-trivial to generate that kind of architecture specifications. However, our experiments at least demonstrate that the LLMs possess the potential to conduct the task of generating architecture specifications. The speed of the LLMs in performing this task, compared to the experienced chip architects, still has significant advantages.

\subsection{Convert RTL code to Architecture Specifications}
\label{subsec:convert}

\begin{figure}[t]
\centering
\includegraphics[width=\linewidth]{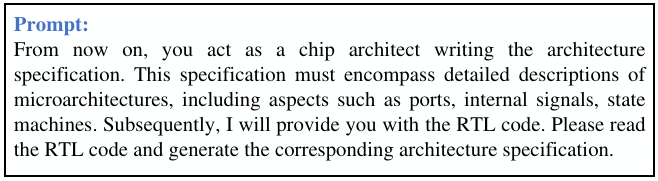}
\caption{An example prompt to initiate writing architecture specifications on the basis of RTL code.  The response is shown in Figure~\ref{fig:detailedWrite2}.}
\vspace{-.2in}
\label{fig:startWrite2}
\end{figure}

\begin{figure}[t]
\centering
\includegraphics[width=\linewidth]{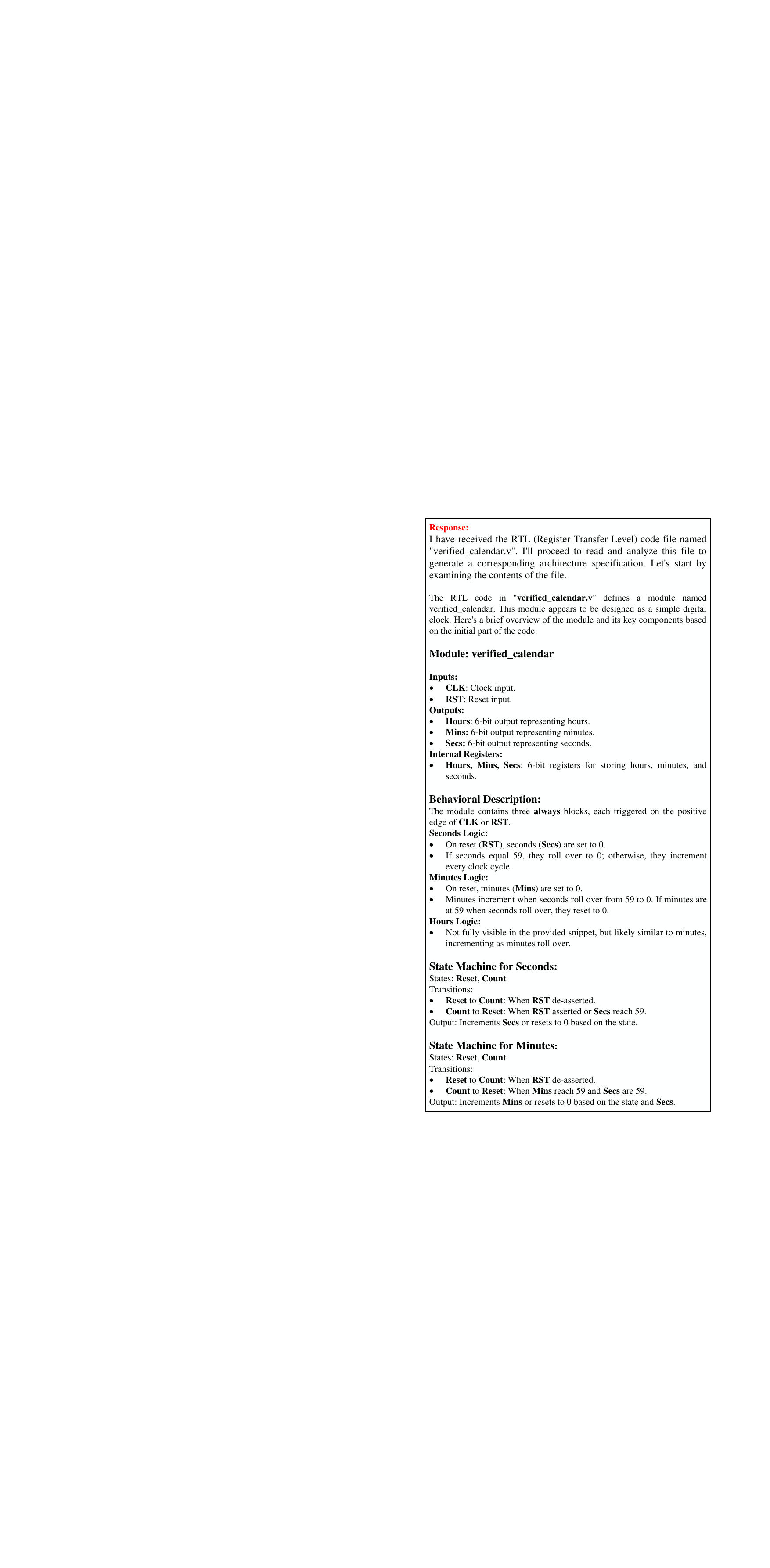}
\caption{A GPT response example to the prompt in Figure~\ref{fig:startWrite2}.}
\vspace{-.2in}
\label{fig:detailedWrite2}
\end{figure}


Figure~\ref{fig:startWrite2} showcases the prompt employed for converting the RTL code into LAS. Following this prompt, we provide the RTL code to the LLM, which can be in the form of textual format or as a source code file. In our example, we adopt the Verilog file that is open-sourced by RTLLM \cite{URL:calendar} and it is uploaded to our GitHub repo. The response of GPT is shown in Figure~\ref{fig:detailedWrite2}. In summary, the generated architecture specifications are slightly shorter than those depicted in Figure~\ref{fig:writeResp}, yet they effectively preserve the correct meaning. We conduct experiments to analyze this phenomenon. Our findings indicate that the provided RTL code confines the scope for the specification generation. Without the RTL code, the LLM may generate the architecture specifications on the basis of the previously trained corpus, which exhibits diverse explanations for the single circuit design. On the contrary, after we offer the LLM with a unique RTL code, its responses are likely to be more orientated.

\section{LLM Reviews Architecture Specification}
\label{sec:review}

\subsection{Motivation}

The accuracy and precision of architecture specifications directly affect the overall quality of chip designs. Ensuring these specifications are correct is crucial for the successful development and functionality of the chips. Traditionally, the chip company should spend efforts to review architecture specifications. Analogous to the reasons mentioned in Section~\ref{subsec:writemoti}, reviewing architecture specifications also requires significant human efforts. LLMs can also help reviewers in these tasks. At least, the reviewers could use these tools to get an overview of the specification files and obtain some comments from them. Fortunately, as we will demonstrate in Section~\ref{subsec:defectsdef} and Section~\ref{subsec:reviewdemo}, the LLMs show promise in achieving more profound objectives in the review of architectural specifications, going beyond basic overviews to provide detailed and meaningful insights.

\subsection{Defects Category}
\label{subsec:defectsdef}

\begin{table}[h]
\small
\centering
 \renewcommand{\arraystretch}{1.3}
\begin{tabular}{|p{0.15\linewidth}|p{0.75\linewidth}|} 
 \hline
 \textbf{Type} & \textbf{Potential Defects} \\  \hline

 Common Case
 & \mydot Typographical Error \newline
 \mydot Inconsistence or Contradiction Error \newline
 \mydot Incomplete or Unclear Error \\ \hline

 Level Specific
 & \mydot Combinational Loops Error (LAS) \newline
 \mydot Uninitialized Register Value Error (LAS) \newline
 \mydot Improvement for Micro-architectural Design (LAS) \newline
 \mydot Improvement for Architectural Design (MAS) \newline
 \\ \hline
 
 Level Spanning
 & \mydot Inconsistence or Contradiction Error \newline (Across Various Levels) \newline \\
    
 \hline
\end{tabular}
\caption{Potential Defects Occurred in Different Levels Architecture Specifications}
\label{tab:spec_defects}
\vspace{-2mm}
\end{table}

Initially, we have pinpointed various types of potential defects that may occur in architecture specifications. Table~\ref{tab:spec_defects} summarizes these defects, categorizing them according to the level of architecture specifications. 

In many instances, there are defects that could manifest in all three tiers of architecture specifications: HAS, MAS, and LAS. They include the \textit{Typographical Error}, \textit{Inconsistence or Contradiction Error}, and \textit{Incomplete or Unclear Error}. The \textit{Inconsistence or Contradiction Error} denotes situations within a single specification file where either two concepts describing the same object are inconsistent, or two related concepts are contradictory. The \textit{Incomplete or Unclear Error} refers to instances where certain concepts lack essential information, resulting in sentences that are open to ambiguous interpretations. 

Furthermore, the potential defects could also be level-specific, indicating that HAS, MAS, or LAS may each have distinct defects. For example, LAS might include details about port connections between various modules. LLMs can be employed to scrutinize the LAS for identifying potential \textit{Combinational Loops Error}. Such scrutiny, however, may not be applicable to HAS and MAS. In addition to locating the writing issues, we also identify some high-level objectives for the review of architecture specification, such as the \textit{Improvement for Architectural Design} and \textit{Improvement for Micro-architectural Design}. These objectives are aimed at leveraging the potential of the review process to enhance both the architectural and micro-architectural design aspects. This goes beyond mere error correction, focusing on the overall optimization and refinement of the design.

Additionally, these defects might span across different levels. One example of this is the \textit{Inconsistence or Contradiction Error (Across Various Levels)}, that the lower-level architecture specifications should not contradict the contents established in the higher-level specifications. Note that it is not equivalent to the \textit{Inconsistence or Contradiction Error} established for the common case. The former indicates the inconsistence or contradiction between different levels of architecture specification files, while the latter confines the error in one individual specification file.

\subsection{Architecture Specification Review}
\label{subsec:reviewdemo}

We exploit the potential defects in architecture specifications as the types of feedback (Section~\ref{subsec:defectsdef}) we aim to receive from the review conducted by the LLMs. We have developed two methodologies to guide the LLMs in reviewing architecture specifications. The first method involves directly uploading the entire specification file and requesting the LLMs to provide feedback section by section. The second method entails splitting the specification file into sections, uploading each section individually, and then asking the LLMs to conduct a review immediately upon the upload of each section. Regarding the first approach, it has the advantage of not requiring extra steps to divide the specification file. However, we've observed that even when instructed to give feedback for each section, the LLMs struggle to accurately recognize the boundaries between different sections. This often results in the provision of incorrect or misleading indication messages by the LLMs. The second approach, which involves splitting the specification file into sections and uploading each one separately for review, can effectively alleviate this problem by providing clearer boundaries for the LLMs to analyze and respond to each section individually. In the subsequent sections, we will use part of the architecture specifications, as shown in Table~\ref{tab:allSpec}, to demonstrate LLM's strengths and weaknesses in reviewing architecture specifications. 

\subsubsection{Reviewing One File at a Time}

\begin{figure}[t!]
\centering
\includegraphics[width=\linewidth]{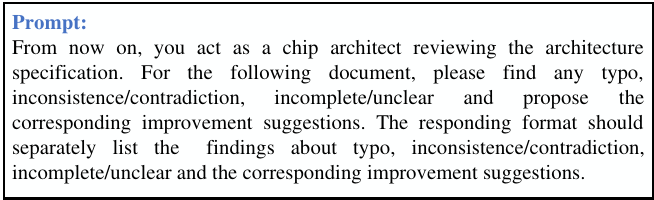}
\caption{An example prompt to initiate the review of architecture specifications. The response is shown in Figure~\ref{fig:startResp}.}
\vspace{-.1in}
\label{fig:startReview}
\end{figure}

\begin{figure}[t!]
\centering
\includegraphics[width=\linewidth]{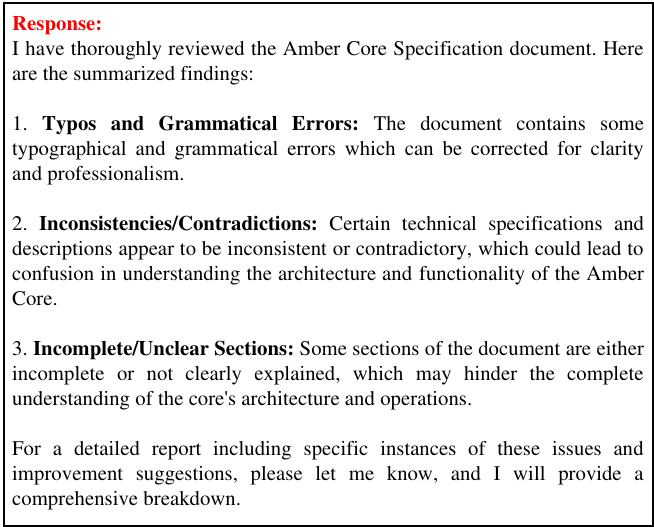}
\caption{A GPT response example to the prompt in Figure~\ref{fig:startReview}.}
\vspace{-.1in}
\label{fig:startResp}
\end{figure}

\begin{figure}[t!]
\centering
\includegraphics[width=\linewidth]{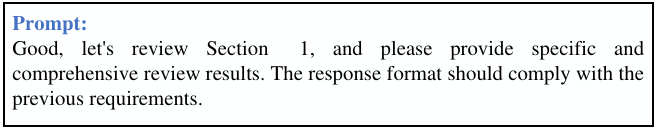}
\caption{An example prompt to review individual sections of the specification document. The response is shown in Figure~\ref{fig:detailedResp}.}
\vspace{-.1in}
\label{fig:detailedReview}
\end{figure}

\begin{figure}[t!]
\centering
\includegraphics[width=\linewidth]{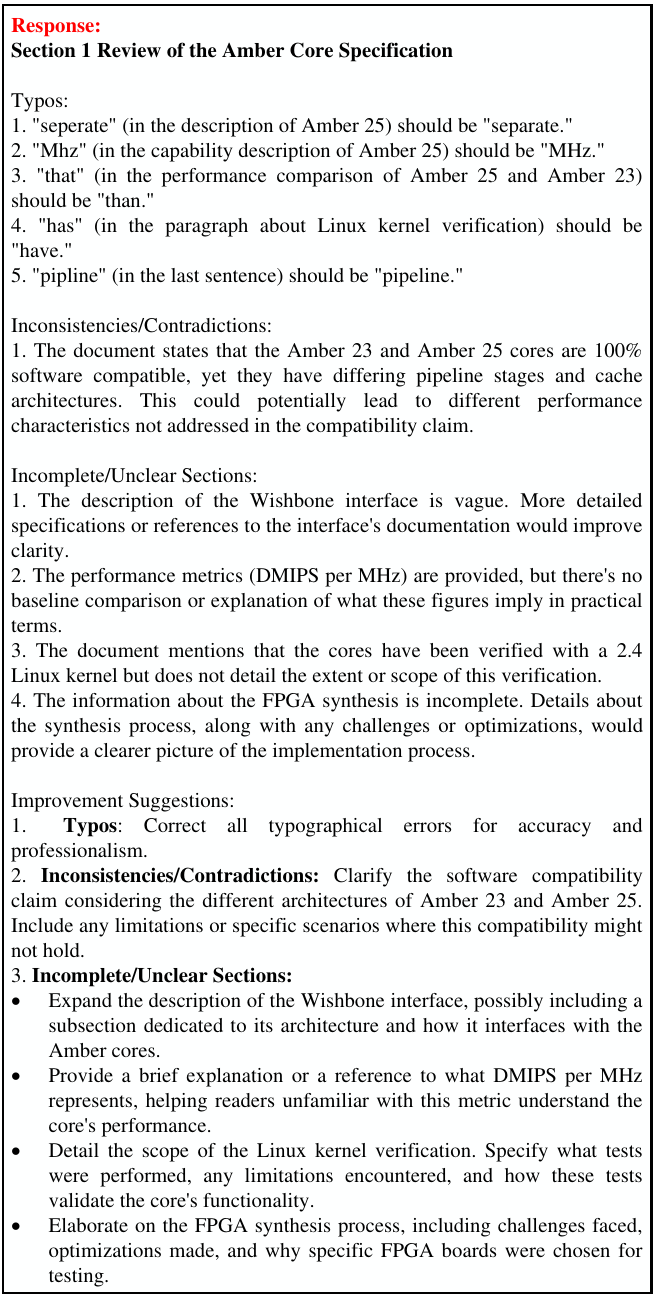}
\caption{A GPT response example to the prompt in Figure~\ref{fig:detailedReview}.}
\vspace{-.2in}
\label{fig:detailedResp}
\end{figure}

We employ the Amber 2 Core Specification \cite{URL:amber} to conduct the experiments of reviewing architecture specifications. The tested file used in our example is from \cite{URL:amber_pdf} and it has been uploaded to our GitHub repo. Its concise contents enable current LLMs like GPT-4 to review it at a time. According to Table~\ref{tab:spec_defects}, we develop a prompt to request GPT-4 providing feedback on common defects. We use the prompt, as illustrated in Figure~\ref{fig:startReview}, to initiate the process of writing architecture specifications. Next, we upload the specification file selected for testing and ask GPT to summarize its findings. As illustrated in Figure~\ref{fig:startResp}, the LLM will then provide an overview of its analysis. This information can offer designers a quick glimpse into the results of the review. Subsequently, as depicted in Figure~\ref{fig:detailedReview}, we request GPT to provide us with detailed review results for each individual section of the specification document. 
Figure~\ref{fig:detailedResp} displays the in-depth review results specifically for Section 1 of the Amber Core Specification. In general, we have observed that GPT-4 is adept at correctly identifying typographical errors. Furthermore, in more complex cases such as inconsistence or contradiction errors, as well as incomplete or unclear errors, GPT-4 is also capable of offering constructive advice. It can give the reasons why these types of errors exist and suggest ways to improve them. Nonetheless, our research has revealed that GPT-4 faces challenges in effectively dividing an architecture specification file into distinct sections. This limitation often results in the feedback provided for one section possibly originating from other sections of the document. To mitigate this problem, we propose to split the complete specification file into individual sections, and request GPT to review them sequentially.

\subsubsection{Splitting and Reviewing File Section-by-Section}

\begin{figure}[t!]
\centering
\includegraphics[width=\linewidth]{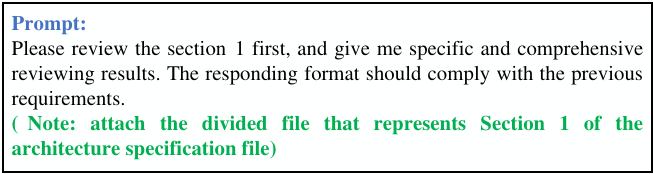}
\caption{An example prompt to review the split files. Though the response from GPT is similar to the Figure~\ref{fig:detailedResp}, the feedback is more likely to concentrate on the individual section.}
\vspace{-.1in}
\label{fig:detailedReview2}
\end{figure}

Similar to the first method, we begin by using prompts, as shown in Figure~\ref{fig:startReview}, to initiate the review of architecture
specifications. Next, instead of uploading the entire architecture specification file for review, we submit only the file containing the contents of the individual section as an attachment. Simultaneously, we provide the prompt, as illustrated in Figure~\ref{fig:detailedReview2}, directing GPT to review each section. In this strategy, each review task for different sections is paired with its corresponding chapter file, ensuring that each section is reviewed in isolation and in context. The responses given by the GPT in this review process maintain a format similar to the outputs generated by the previous, more holistic review method. However, we have observed that the feedbacks are more likely to concentrate on each individual section. Furthermore, the GPT appears more effective in identifying a broader range of errors, which increases the number of available output items. The accurate and abundant review results enable designers to quickly but comprehensively locate the defects in their specification documents.

\subsection{Feedbacks Evaluation}

Not all feedback provided by the LLM is correct or useful. Therefore, it is essential to have strategies in place to evaluate the review results of architecture specifications effectively. The most intuitive method is to ask the designers to check the outputs of the LLM. This approach, compared to directly reviewing the architecture specifications themselves, is already more efficient, potentially saving significant time and effort. Nonetheless, an alternative approach involves training a new language model specifically for evaluating and filtering the review results provided by the LLMs. This specialized model could be designed to assess the relevance and accuracy of the LLM's feedback \cite{chang2023survey}, thereby streamlining the review process and enhancing the reliability of the outcomes. We leave this part for future exploration.

\section{Conclusion}

In this paper, we propose a novel framework for utilizing LLMs to generate and review architecture specifications. We defined architecture specifications, categorized them into clear levels, and compiled a corresponding dataset. Our approach focuses on two primary applications of LLMs: firstly, in generating architecture specifications, including assistance in writing and converting RTL code into specifications, and secondly, in the review of these specifications. This innovative methodology signifies a transformative step in architecture specification development, offering a path toward more efficient, accurate, and streamlined processes in chip design.

\bibliographystyle{IEEEtran}
\bibliography{spec, references_1, references_2}
\end{document}